\documentstyle[12pt]{article}

\newcommand{\be}{\begin{equation}}
\newcommand{\ee}{\end{equation}}

\newcommand{\PRL}{{\it Phys.~Rev.~Lett.~}}
\newcommand{\JMP}{{\it J.~Math.~Phys.~}}
\newcommand{\JHEP}{{\it JHEP~}}
\newcommand{\CMP}{{\it Comm.~Math.~Phys.~}}
\newcommand{\NP}{{\it Nucl.~Phys.~}}
\newcommand{\AP}{{\it Ann.~Phys.~}}
\newcommand{\PL}{{\it Phys.~Lett.~}}

\newcommand{\Tr}{{\rm Tr}}
\newcommand{\tr}{{\rm tr}}
\newcommand{\A}{{\rm A}}
\newcommand{\C}{{\rm C}}
\newcommand{\Ch}{{\hat {\rm C}}}
\newcommand{\F}{{\rm F}}
\newcommand{\Fh}{{\hat {\rm F}}}
\newcommand{\Sh}{{\hat S}}
\newcommand{\dd}{{\rm d}}
\newcommand{\D}{{\rm D}}
\expandafter\ifx\csname mathbbm\endcsname\relax

\else

\fi
\begin{document}
\begin{titlepage}
\begin{flushleft}  
       \hfill                      {\tt hep-th/0010264}\\
       \hfill                      October 2000\\
\end{flushleft}
\vspace*{3mm}
\begin{center}
{\LARGE Noncommutative Chern-Simons terms and the noncommutative vacuum
\\}
\vspace*{12mm}
{\large Alexios P. Polychronakos\footnote{On leave from Theoretical 
Physics Dept., Uppsala University, 751 08 Sweden and Dept.~of Physics, 
University of Ioannina, 45110 Greece; E-mail: poly@teorfys.uu.se} \\
\vspace*{5mm}
{\em Physics Department, City College of the CUNY \\
New York, NY 10031, USA}\\
\vspace*{5mm}
and \\
\vspace*{5mm}
{\em Physics Department, The Rockefeller University \\
New York, NY 10021, USA\/}\\}
\vspace*{15mm}
\end{center}

\begin{abstract}
It is pointed out that the space noncommutativity parameters 
$\theta^{\mu \nu}$ in noncommutative gauge theory can be considered as 
a set of superselection parameters, in analogy with the $\theta$-angle 
in ordinary gauge theories. As such, they do not need to enter explicitly
into the action.  A simple generic formula is then suggested to reproduce
the Chern-Simons action in noncommutative gauge theory, which reduces to 
the standard action in the commutative limit but in general implies a cascade
of lower-dimensional Chern-Simons terms. The presence of these terms in 
general alters the vacuum structure of the theory and nonstandard gauge 
theories can emerge around the new vacua.
\end{abstract}

\end{titlepage}

\section{Introduction}

Chern-Simons actions have appeared in various contexts as
topological terms in the action for gauge fields in odd
(spacetime) dimensions \cite{DJT}. Their densities are essentially defined 
as the gauge-field dependent differential form whose exterior derivative
equals $\tr \F^{n+1}$, with $\F$ the field strength. Specifically,
we define the gauge field one-form and the field strength two-form as
\be
\A = i A_\mu dx^\mu ~,~~~\F = \dd \A + \A^2 = \frac{i}{2} (\partial_\mu 
A_\nu - \partial_\nu A_\mu + i [A_\mu , A_\nu ] ) dx^\mu dx^\nu 
\ee
The Chern-Simons action $S_{2n+1}$ is the integral of the $2n+1$-form 
$\C_{2n+1}$ satisfying
\be
\dd \C_{2n+1} = \tr \F^{n+1}
\label{dCF}
\ee
By virtue of (\ref{dCF}) and the gauge invariance of $\tr \F^n$
it follows that $S_{2n+1}$ is gauge invariant up to total derivatives,
since, if $\delta$ stands for an infinitesimal gauge transformation,
\be
\dd \delta \C_{2n+1} = \delta \dd \C_{2n+1} = \delta \tr \F^n = 0 ~,
~~{\rm so}~~~ \delta C_{2n+1} = \dd \Omega_{2n}
\ee
The integrated action is therefore invariant under infinitesimal
gauge transformations. Large gauge transformations may lead to 
an additive change in the action and they usually imply a quantization
of its coefficient \cite{DJT} (for a review see \cite{OA}).
As a result, the equations of motion derived from
this action are gauge covariant and read
\be
\frac{\delta S_{2n+1}}{\delta \A} = \frac{\delta}{\delta \A} 
\int \C_{2n+1} = (n+1) \F^n 
\label{defC}
\ee
We shall consider (\ref{defC}) as the defining relation for $\C_{2n+1}$.

\section{Noncommutative gauge theory}

Similar actions can be defined in noncommutative theories 
\cite{CF,TK,CW,MS,GS}. These are gauge theories based upon 
noncommutative spaces \cite{CDS,SW}. The coordinates 
$X^\mu$ of such spaces obey the commutation relations
\be
[ X^\mu , X^\nu ] = i \theta^{\mu \nu} ~,~~~\mu,\nu=1, \dots d
\label{XX}
\ee
The antisymmetric two-tensor $\theta^{\mu \nu}$ is usually taken to 
commute with all $X^\mu$ and is, thus, a set of constant c-numbers.
Its inverse
\be
\omega_{\mu \nu} = (\theta^{-1} )_{\mu \nu}
\ee
defines a constant two-form $\omega$ characterising the 
noncommutativity of the space. Clearly $\theta$ and $\omega$ can be
brought into a canonical (Darboux) form by an appropriate orthogonal
rotation of the space, breaking it into noncommutative two-dimensional
subspaces. Some coordinates may, then, be commuting (odd-dimensional 
spaces necessarily have at least one commuting coordinate). In general,
there will be $2n$ properly noncommuting coordinates $X^\alpha$ ($\alpha
= 1, \dots 2n$) and $q=d-2n$ commuting ones $Y^i$ ($i=1,\dots q$). 
In that case $\omega$ will be defined 
as the inverse of the projection $\bar \theta$ of $\theta$ on the fully
noncommuting subspace:
\be
\omega_{\alpha \beta} = ({\bar \theta}^{-1} )_{\alpha \beta} ~,~~~
\omega_{ij} = 0
\ee

The actual spacetime can be though of as a representation of the above
operator algebra (\ref{XX}). For the commuting components $Y^i$ this 
representation must necessarily be reducible (else the corresponding
directions would effectively be absent, consisting of a single point); 
states are labeled by the values of these coordinates $y^i$, taken to be 
continuous. The rest of the space is
represented by the tensor product of Heisenberg-Fock space representations
(one for each two-dimensional noncommuting subspace $k=1, \dots n$).
In general, we can have the direct sum of $N$ such irreducible components
for each set of values $y^i$, labeled by an extra index $a=1,\dots N$
(we shall take $N$ not to depend on $y^i$). A complete basis for the states,
then, can be
\be
|n_1 , \dots n_n ; y^1 , \dots y^q ; a>
\ee
where $n_k$ is the Fock (oscillator) excitation number of the $k$-th 
two-dimensional subspace.

Due to the reducibility of the above representation, the operators
$X^\mu$ do not constitute a complete set.
An important set of additional operators are translation (derivative)
operators $\partial_\mu$. These are defined through their
action on $X^\mu$, generating constant shifts:
\be
[\partial_\mu , X^\nu ] = \delta_\mu^\nu
\ee
On the fully noncommutative subspace these are inner automorphisms
generated by
\be
\partial_\alpha = -i \omega_{\alpha \beta} X^\beta
\ee
For the commutative coordinates, however, extra operators have to be
appended, shifting the Casimirs $Y^i$ and thus acting on the 
coordinates $y^i$ as usual derivatives. There is yet another set of operators
in the full representation space: $G^r$, $r=1, \dots N^2$,  the hermitian operators
mixing the irreducible components $a$. The set of $X^\alpha , \partial_i , G^r$ is
now complete.

The integral over space is defined as the trace in the full representation,
normalized as:
\be
\int d^d x = \int d^q y  ~ \tr'  ~\sqrt{\det (2\pi\theta )} \, \tr \equiv \Tr  
\ee
where $\tr$ is the trace over the Fock spaces and $\tr'$ is the trace
over the degeneracy index $a=1,\dots N$.

Gauge fields $A_\mu$ are arbitrary functions of the noncommutative 
coordinates $X^\mu$ and thus are (hermitian) operators acting on 
the full representation space. Since they do not depend on $\partial_i$
they cannot shift the values of $y^i$, while they act nontrivially on the
fully noncommuting subspace. They have effectively become big matrices
acting on the full Fock space with elements depending on the commuting
coordinates. Derivatives of these fields are defined through the
adjoint action of $\partial_\mu$
\be
\partial_\mu \cdot A_\nu = [ \partial_\mu , A_\nu ]
\ee
Using the above formalism, gauge field theory can be built in a way
analogous to the commuting case. Gauge transformations are unitary
transformations in the full representation space. Restricting $A_\mu$ to 
depend on the coordinates only, as above, produces the so-called $U(1)$
gauge theory. $U(N)$ gauge theory can be obtained by relaxing this
restriction and allowing $A_\mu$ to also be a function of the $G^r$ and
thus act on the index $a$ (see also \cite{GN}). Further generalizations in which
$A_\mu$ are also allowed to depend on $\partial_i$ and act nontrivially
on $y^i$, thus becoming completely general hermitian operators, are possible
but are usually not considered.

We should mention that an alternative way to describe field theories on 
noncommutative spaces is by  ordering the operators $X_\mu$ in the expressions for 
the fields in a specific way and thus establishing a one-to-one  correspondence
between functions of operators and ordinary functions. (The Weyl symmetric
ordering is usually adopted.) The product of operators, then,
is mapped to the so-called star-product of functions, while derivatives and integrals
are the standard commutative ones. This approach has the advantage of bypassing
some of the conceptual issues arising in noncommutative theories, but also the
disadvatntage of obscuring their basic simplicity and some of their interesting structure.
We shall employ the operator formalism in this paper.

\section{Superselection of the noncommutative vacuum}

The basic moral of the above discussion is that noncommutative gauge theory
can be written in a universal way. In the operator formulation no special distinction 
needs be done between $U(1)$ and $U(N)$ theories, nor need gauge and spacetime
degrees of freedom be treated distinctly. The fundamental operators of the theory are
\be
D_\mu = -i\partial_\mu + A_\mu
\ee
corresponding to covariant derivatives. Under gauge transformations they transform
covariantly: 
\be
D_\mu \to U D_\mu U^{-1}
\ee
Any lagrangian built entirely out of $D_\mu$ will lead to a gauge invariant action, 
since the trace will remain invariant under any unitary transformation. The standard
Lorentz-Yang-Mills action is built by defining the field strength
\be
F_{\mu \nu} = \partial_\mu \cdot A_\nu - \partial_\nu \cdot A_\mu + i
[ A_\mu , A_\nu ] = i [ D_\mu , D_\nu ] - \omega_{\mu \nu}
\ee
and writing the standard action
\be
S_{LYM} = \frac{1}{4g^2} \Tr F_{\mu \nu} F^{\mu \nu} = - \frac{1}{4g^2} \Tr
 ([ D_\mu , D_\nu ] +i \omega_{\mu \nu} )^2
\ee
(we used some metric tensor $g^{\mu \nu}$ to raise the indices of $F$). Note that
the operators $\partial_\alpha \cdot$, understood to act in the adjoint on fields, commute,
while the operators $\partial_\alpha = -i \omega_{\alpha \beta} X^\beta$ have a nonzero
commutator equal to 
\be
[\partial_\alpha , \partial_\beta ] = i \omega_{\alpha \beta}
\ee
This explains the extra $\omega$-term appearing
in the definition of $F$ through covariant derivative commutators.

As was pointed out in \cite{AP}, however, one can just as well work with the action
\be
{\hat S}_{LYM} = \frac{1}{4g^2} \Tr {\hat F}_{\mu \nu} {\hat F}^{\mu \nu} = 
- \frac{1}{4g^2} \Tr [ D_\mu , D_\nu ]^2
\label{DDD}
\ee
Indeed, $\hat S$ differs from $S$ by a term proportional to $\Tr \omega^2$, which is an
irrelevant (infinite) constant, as well as a term proportional to $\omega^{\mu \nu}
\Tr [ D_\mu , D_\nu ]$, which, being the trace of a commutator (a `total derivative'),
does not contribute to the equations of motion. The two actions then lead to the same
classical theory. The equations of motion for the operators $D_\mu$ are
\be
[D_\mu , [ D_\mu , D_\nu ]] = 0
\ee
Apart from the trivial solution $D_\mu = 0$ this has as solution all 
operators with c-number commutators, satisfying
\be
[ D_\mu , D_\nu ] = -i\omega_{\mu \nu}
\ee
for some $\omega$. This is the classical `noncommutative vacuum', where 
$D_\mu = -i\partial_\mu$, and expanding $D_\mu$ around this vacuum leads 
to noncommutative gauge theory. Quantum mechanically $\omega_{\mu \nu}$ are 
superselection parameters and the above vacuum is stable.  To see this, 
assume that the time direction is commutative and consider the collective mode
\be
D_\alpha = -i\lambda_{\alpha \beta} \partial_\beta
\ee
with $\lambda_{\alpha \beta}$ parameters depending only on time.
This mode would change the noncommutative vacuum while leaving the gauge
field part of $D_\alpha$ unexcited. $\omega$ gets modified into
\be
\omega_{\mu \nu}' = \lambda_{\mu \alpha} \omega_{\alpha \beta} 
\lambda_{\beta \nu}
\ee
The action implies a quartic potential for this mode, with a strength proportional
to $\Tr 1$, and a kinetic term proportional to $\Tr \partial_\alpha \partial_\beta$.
(There is also a gauge constraint which does not alter the qualitative dynamical
behavior of $\lambda$.) Both potential and kinetic terms are infinite, 
and to regularize them we should truncate each Fock space
trace up to some highest state $\Lambda$, corresponding to a finite volume 
regularization (the area of each noncommutative two-dimensional subspace has 
effectively become $\Lambda$). It can be seen that the potential term would 
grow as $\Lambda^n$ while the kinetic term
would grow as $\Lambda^{n+1}$. Thus the kinetic term dominates; the above 
collective degrees of freedom acquire
an infinite mass and will remain ``frozen'' to whatever initial value
they are placed, in spite of the nontrivial potential. (This is analogous to
the $\theta$-angle of the vacuum of four-dimensional nonabelian gauge theories:
the vacuum energy depends on $\theta$ which is still superselected.)
Quantum mechanically there is no interference between different
values of $\lambda$ and we can fix them to some c-number value, thus fixing
the noncommutativity of space. This phenomenon is similar to symmetry breaking,
but with the important difference that the potential is not flat along changes of the
``broken'' vacuum, and consequently there are no Goldstone bosons.

In conclusion, we can start with the action (\ref{DDD}) as the definition of our
theory, where $D_\mu$ are arbitrary operators (matrices) in some space. Gauge
theory is then defined as a perturbation around a (stable) classical vacuum.
Particular choices of this vacuum will lead to standard noncommutative gauge
theory, with $\theta^{\mu \nu}$ and $N$ appearing as vacuum parameters.

\section{Noncommutative Chern-Simons action}

We are now set to define Chern-Simons actions. To this end, we shall define the
usual basis of one-forms $dx^\mu$ as a set of formal anticommuting parameters
with the property 
\be
dx^\mu dx^\nu = - dx^\nu dx^\mu ~,~~~
dx^{\mu_1} \cdots dx^{\mu_d} = \epsilon^{\mu_1 \dots \mu_D}
\ee
Topological actions do not involve the metric tensor and can be written as 
integrals of $d$-forms. The only dynamical objects available in our theory 
are $D_\mu$ and thus the only form that we can write is
\be
\D = i dx^\mu D_\mu = \dd + \A
\ee
where we defined the exterior derivative and gauge field one-forms
\be
\dd = dx^\mu \partial_\mu ~,~~~ \A = i dx^\mu A_\mu 
\ee
(note that both $\D$ and $\A$ are antihermitian). 
The action of the exterior derivative $\dd$ on an operator $p$-form $H$,
$\dd \cdot {\rm H}$, 
yields the $p+1$-form $dx^\mu [\partial_\mu , {\rm H}]$ and is given by
\be
\dd \cdot {\rm H} = \dd {\rm H} - (-)^p {\rm H} \dd
\ee
In particular, on the gauge field one-form $\A$ it acts as 
\be
\dd \cdot \A = \dd \A + \A \dd
\ee
Correspondingly, the covariant exterior derivative of $\rm H$ is
\be
\D \cdot {\rm H} = \D {\rm H} - (-)^p {\rm H} \D
\ee
As a result of the noncommutativity of the operators $\partial_\mu$,
the exterior derivative operator is not nilpotent but rather satisfies
\be
\dd^2 = \omega ~,~~~~ \omega = \frac{i}{2} dx^\mu dx^\nu \omega_{\mu \nu}
\ee
We stress, however, that $\dd \cdot$ is still nilpotent since $\omega$
commutes with all operator forms:
\be
\dd \cdot \dd \cdot {\rm H} = [\dd , [\dd , {\rm H}]_\mp ]_\pm =
\pm [\omega , {\rm H} ] = 0
\ee
The two-form $\Fh = \frac{i}{2}dx^\mu dx^\nu {\hat F}_{\mu \nu}$ is simply
\be
\Fh = \D^2 = \frac{1}{2} \D \cdot \D = 
\omega + \dd \A + \A \dd + \A^2 = \omega + \F
\ee
where $\F = \frac{i}{2} dx^\mu dx^\nu F_{\mu \nu}$ is the conventionally
defined field strength two-form.

The most general $d$-form that we can write involves arbitrary combinations of
$\D$ and $\omega$. If, however, we adopt the view that $\omega$ should arise as
a superselection (vacuum) parameter and not as a term in the action, the unique
form that we can write is $\D^d$ and the unique action
\be
\Sh_d = \frac{d+1}{2d} \Tr \, \D^d = \Tr \, \C_d
\label{SCS}
\ee
This is the Chern-Simons action. The coefficient was chosen to conform with
the commutative definition (\ref{defC}), as will be seen shortly. In even dimensions 
$\Sh_d$ reduces to 
the trace of a commutator $\Tr [\D , \D^{d-1} ]$, a total derivative that 
does not affect the equations of motion and corresponds to a topological term.
In odd dimensions it becomes a nontrivial action.

$\Sh_d$ is by construction gauge invariant. To see that it also 
satisfies the defining property of a Chern-Simons form (\ref{defC}) 
is almost immediate: $\delta / \delta \A = \delta / \delta \D$ and thus,
for $d=2n+1$:
\be
\frac{\delta}{\delta \A} \Tr \, \D^{2n+1} = (2n+1) \, \D^{2n} = (2n+1) \, \Fh^n
\ee
So, with the chosen normalization in (\ref{SCS}) we have the defining condition
(\ref{defC}) with $\Fh$ in the place of $\F$. What is less obvious is that $\Sh_D$
can be written entirely in terms of  $\F$ and $\A$ and that, for commutative spaces,
it reduces to the standard Chern-Simons action. To achieve that, one must expand
$\C_D$ in terms of $\dd$ and $\A$, make use of the cyclicity of trace and the condition
$\dd^2 = \omega$ and reduce the expressions into ones containing 
$\dd \A + \A \dd$ 
rather than isolated $\dd$'s.  The condition 
\be
\Tr \omega^n \dd = 0 
\ee
which is a result of the fact that $\partial_\mu$ is off-diagonal for both 
commuting and noncommuting dimensions, can also be used to get rid of overall 
constants. This
is a rather involved procedure for which we have no algorithmic approach.
In the appendix we will illustrate the case $d=7$, which presents the full range of
complexity in terms of reduction to ordinary terms. Note,
further, that the use of the cyclicity of trace implies that we dismiss total derivative
terms (traces of commutators). Such terms do not affect the equations of motion.
For $d=1$ the result is simply
\be
\Sh_1 = \Tr \A
\ee
which is the `abelian' one-dimensional Chern-Simons term. 
For $d=3$ we obtain
\be
\Sh_3 = \Tr (\A \F - \frac{1}{3} \A^3 ) + 2 \Tr (\omega \A )
\ee
where we used the fact that $\Tr [\A (\dd \A + \A \dd )] = 2\Tr (\A^2 \dd)$.  The first
term is the noncommutative version of the standard three-dimensional 
Chern-Simons term, while the second is a lower-dimensional Chern-Simons 
term involving explicitly $\omega$.

We can get the general expression for $\Sh_d$ by referring to the defining
relation. This reads
\be
\frac{\delta}{\delta \A} \Sh_{2n+1} = (n+1) \Fh^n = (n+1) (\F +\omega)^n
= (n+1) \sum_{k=0}^n {n \choose k} \omega^{n-k} \F^k 
\ee
and by expressing $\F^k$ as the $\A$-derivative of the standard Chern-Simons
action $S_{2k+1}$ we get
\be
\frac{\delta}{\delta \A}  \left\{ \Sh_{2n+1}  - \sum_{k=0}^n 
{n+1 \choose k+1}  \omega^{n-k}  S_{2k+1} \right\} = 0
\ee
So the expression in brackets must be a constant, easily seen to be zero by
setting $\A=0$. We therefore have
\be
\Sh_{2n+1}  = \sum_{k=0}^n {n+1 \choose k+1}  \Tr \omega^{n-k}  \C_{2k+1} 
\label{SSh}
\ee
We observe that we get the $2n+1$-dimensional Chern-Simons action plus
all lower-dimensional actions with tensors $\omega$ inserted to complete the
dimensions. Each term is separately gauge invariant and we could have chosen
to omit them, or include them with different coefficient. It is the specific combination
above, however, that has the property that it can be reformulated in a way that does
not involve $\omega$ explicitly. The standard Chern-Simons action can also
be written in terms of $\D$ alone by inverting (\ref{SSh}):
\be
S_{2n+1} = (n+1) \Tr \int_0^1 (t^2 \D^2 -\omega)^n dt =  \Tr \sum_{k=0}^n 
{n+1 \choose k+1} \frac{k+1}{2k+1}  (-\omega)^{n-k} \D^{2k+1}
\label{ShS}
\ee

We also point out a peculiar propery of the Chern-Simons form $\Ch_{2n+1}$.
Its covariant derivative yields $\Fh^{n+1}$:
\be
\D \cdot \Ch_{2n+1} = \D \Ch_{2n+1} + \Ch_{2n+1} \D = \frac{2n+2}{2n+1}
\, \Fh^{n+1}
\label{DCF}
\ee
A similar relation holds between $C_d$ (understood as the form appearing inside the
trace in the right hand side of (\ref{ShS})) and $\F$. Clearly the standard Chern-Simons 
form does not share this property.  Our $C_d$
differs from the standard one by commutators that cannot all be written as ordinary
derivatives (such as, {\it e.g.}, $[\dd , \dd \A]$). These unconventional terms turn
$C_d$ into a covariant quantity that satisfies (\ref{DCF}).

We remark here that a construction reminiscent of the above exists in string
field theory \cite{W}, where the action is written as the integral of $(Q+A)^3$, with 
$Q$ the BRST charge. We also remark that actions similar to the subleading terms in
(\ref{SSh}) were examined a while ago in the commutative case \cite{NS}. 
In particular, the five-dimensional action $\omega S_3$, with $\omega$ a K\"ahler
form, was considered as a way to construct conformally invariant theories in
four dimensions.

\section{New vacua}

We conclude with the observation that the addition of the Chern-Simons action 
changes the equations of motion of the theory and thus it modifies the noncommutative
vacuum. As an example, in three dimensions the equations of motion read
\be
[ D_\nu , [ D_\nu, D_\mu ]] = i \lambda \epsilon_{\mu \nu \rho} [ D_\nu , D_\rho ]
\ee
with $\lambda$ a coupling constant. (A similar equation was obtained from
a different point of view in \cite{ARS}.) For euclidean metric this has 
finite dimensional solutions. Specifically, 
\be
D_\mu = -\lambda J_\mu ~,~~ J^2 = j(j+1)
\ee
where $J_\mu$ are three $SU(2)$ matrices in an arbitrary spin representation.
Correspondingly, for Minkowski metric we get as solutions all the 
(infinite-dimensional) unitary representations of $SU(1,1)$. The resulting
gauge theories deserve further investigation.

\vskip 0.2in
{\it Acknowledgement}: I would like to thank V.P.~Nair for interesting comments
and discussions.
\vskip 0.3in
{\centerline {\bf Appendix}}
\vskip 0.2in

We shall calculate $\Sh_7$ starting from the definition 
\be
\Sh_7 = \frac{4}{7} \Tr \D^7 = \frac{4}{7} \Tr (\dd + \A )^7
\ee
Expanding $(\dd + \A )^7$ inside the trace we will get all combinations of $\dd$
and $\A$ positioned in a cyclic arrangement. Clearly if two $\dd$ are adjacent on
the cycle they will contract to $\omega$ and will contribute towards lower-order
Chern-Simons terms, of order $\omega$ or more. Concentrating on the leading term
of order $\omega^0$, the relevant combinations are
\be
\Sh_7 = \frac{4}{7}\Tr ( \A^7 + 7 \A^6 \dd + 7 \A^4 \dd \A \dd 
+ 7 \A^3 \dd \A^2 \dd + 7 \A^2 \dd \A \dd \A \dd ) + {\cal O} (\omega )
\ee
From now on all relations will be understood as holding to order $\omega^0$,
omitting terms of order $\omega$ without explicitly writing $+ {\cal O} (\omega )$.

The second term can be expressed as
\be
\Tr (\A^6 \dd ) = \frac{1}{2} \Tr [ \A^5 (\dd \A + \A \dd )] = \frac{1}{2} \Tr [\A^5 (\F - \A^2)]
\ee
To reduce the next two terms we observe
\be
(\F - \A^2 )^2 = (\dd \A + \A \dd )^2 = \dd \A \dd \A + \dd \A^2 \dd + 
\A \dd \A \dd
\ee
and
\be
(\F - \A^2 ) \A (\F - \A^2) = (\dd \A + \A \dd ) \A (\dd \A + \A \dd ) =
\dd \A^2 \dd \A + \dd \A^3 \dd + \A \dd \A \dd \A + \A \dd \A^2 \dd
\ee
Tracing the previous two relations with $\A^3$ and $\A^2$ respectively we get
\be
\Tr [ \A^3 (\F - \A^2)^2 ] = 2\Tr (\A^4 \dd \A \dd ) + \Tr (\A^3 \dd \A^2 \dd )
\ee
\be
\Tr [ \A^2 (\F -\A^2 ) \A (\F - \A^2 )] = 3\Tr (\A^3 \dd \A^2 \dd )+ \Tr (\A^4 \dd \A \dd )
\ee
and solving for the two terms in the right-hand side we obtain
\be
\Tr (\A^4 \dd \A \dd ) = \frac{3}{5} \Tr [ \A^3 (\F - \A^2)^2 ]  - \frac{1}{5}
\Tr [ \A^2 (\F -\A^2 ) \A (\F - \A^2 )]
\ee
\be
\Tr (\A^3 \dd \A^2 \dd ) = \frac{2}{5}  \Tr [ \A^2 (\F -\A^2 ) \A (\F - \A^2 )]
-\frac{1}{5} \Tr [ \A^3 (\F - \A^2)^2 ] 
\ee
Finally, to obtain the last term we observe
\be
(\F - \A^2 )^3 = (\dd \A + \A \dd )^3 = \dd \A \dd \A \dd \A + \dd \A \dd \A^2 \dd
+ \dd \A^2 \dd \A \dd \A + \A \dd \A \dd \A \dd
\ee
and tracing with $\A$
\be
\Tr (\A^2 \dd \A \dd \A \dd ) = \frac{1}{4} \Tr [ \A (\F - \A^2 )^3 ]
\ee
Putting together all the terms we obtain $\Sh_7 = S_7 + {\cal O} (\omega )$ as
\be
S_7 = \Tr \left( \A \F^3 - \frac{2}{5} \A^3 \F^2  
- \frac{1}{5} \A^2 \F \A \F +\frac{1}{5} \A^5 \F -\frac{1}{35} \A^7 \right)
\ee
which is the standard 7-dimensional Chern-Simons action.


\begin{thebibliography}{99}

\bibitem{DJT} S.~Deser, R.~Jackiw and S.~Templeton, \PRL {\bf 48} (1982)
975 and \AP {\bf140} (1982) 372.

\bibitem{OA} O.~Alvarez, \CMP {\bf100} (1985) 279.

\bibitem{CF} A.H.~Chamseddine and J.~Fr\"ohlich, 
\JMP {\bf 35} (1994) 5195

\bibitem{TK} T.~Krajewski, math-phys/9810015

\bibitem{CW} G.-H.~Chen and Y.-S. Wu, hep-th/0006114

\bibitem{MS} S.~Mukhi and N.V.~Suryanarayana, hep-th/0009101

\bibitem{GS} N.~Grandi and G.A.~Silva, hep-th/0010113

\bibitem{CDS} A.~Connes, M.~Douglas and A.~Schwartz, \JHEP
{\bf 9802} (1998) 003

\bibitem{SW} N.~Seiberg and E.~Witten, \JHEP {\bf 9909} (1999) 032

\bibitem{GN} D.J.~Gross and N.A.~Nekrasov, hep-th/0010090

\bibitem{AP} A.P.~Polychronakos, hep-th/0007043

\bibitem{W} E.~Witten, \NP {\bf B268} (1986) 253

\bibitem{NS} V.P~Nair and J.~Schiff, \PL {\bf B246} (1990) 423 and
\NP {\bf B371} (1992) 329

\bibitem{ARS} A.Yu.~Alekseev, A. Recknagel and V.~Schomerus, \JHEP
{\bf 0005} (2000) 010

\end{thebibliography}
\end{document}